\documentclass[conference,a4paper]{IEEEtran}

\usepackage[cmex10]{amsmath}
\usepackage{amssymb}

\usepackage{cases}
\usepackage{multirow}
\usepackage{empheq}

\usepackage{cite}
\usepackage{verbatim}
\usepackage{epsfig}

\usepackage{color}

\newtheorem{theorem}{Theorem}

\DeclareMathOperator{\rank}{rank}

\begin{document}

\sloppy

\title{Redundancy Allocation of Partitioned Linear Block Codes}

\author{
  \IEEEauthorblockN{Yongjune Kim and B. V. K. Vijaya Kumar}
  \IEEEauthorblockA{Dep. of Electrical \& Computer Eng., Data Storage Systems Center (DSSC)\\
    Carnegie Mellon University\\
    Pittsburgh, PA, USA\\
    Email: yongjunekim@cmu.edu, kumar@ece.cmu.edu}
}



\maketitle

\begin{abstract}
Most memories suffer from both permanent defects and intermittent random errors. The partitioned linear block codes (PLBC) were proposed by Heegard to efficiently mask stuck-at defects and correct random errors. The PLBC have two separate redundancy parts for defects and random errors. In this paper, we investigate the allocation of redundancy between these two parts. The optimal redundancy allocation will be investigated using simulations and the simulation results show that the PLBC can significantly reduce the probability of decoding failure in memory with defects. In addition, we will derive the upper bound on the probability of decoding failure of PLBC and estimate the optimal redundancy allocation using this upper bound. The estimated redundancy allocation matches the optimal redundancy allocation well.
\end{abstract}

\section{Introduction}

Most memory systems (e.g., flash memory, phase-change memory, etc.) exhibit two types of imperfections that threaten the data reliability. The first type is a defective memory cell, i.e., defect, whose cell value is stuck-at a particular value independent of the input. For example, some of the cells of a binary memory may be stuck-at 0, and when a 1 is attempted to be written into a stuck-at 0 cell, an error results. The second type of imperfection is a noisy cell which can occasionally result in a random error. The distinction between these two types of imperfections is that defects are permanent, whereas random errors caused by noise are intermittent. Often the terms hard and soft errors are used to describe stuck-at errors of defects and noise-induced random errors, respectively \cite{kuznetsov1974coding, kuznetsov1978, heegard1983capacity, heegard1983plbc}.

By carefully testing the memory, it is possible to know the defect information such as locations and stuck-at values, and this information can be exploited in the encoder and/or the decoder for more efficient coding. This problem was first addressed by Kuznetsov and Tsybakov \cite{kuznetsov1974coding}. They assumed that the locations and stuck-at values of the defects are available to the encoder, but not to the decoder \cite{kuznetsov1974coding, kuznetsov1978}.

Later, Heegard proposed the partitioned linear block codes (PLBC) that efficiently incorporate the defect information in the encoding process and are capable of correcting both stuck-at errors (due to defects) and random errors \cite{heegard1983plbc}. Recently, his work has drawn attention for nonvolatile memories because flash memories and phase change memories (PCM) suffer from defects as well as random errors \cite{lastras2010algorithm, Hwang2011}.

The PLBC require two generator matrices. One of them is for correcting stuck-at errors by masking defects. Masking defects is to find a codeword whose values at the locations of defects match the stuck-at values at those locations \cite{kuznetsov1978, heegard1983plbc}. The other generator matrix is for correcting random errors, which is same as a generator matrix of standard error control coding.

Since the PLBC have two generator matrices, we can separate the redundancy for masking defects from the redundancy for correcting random errors \cite{heegard1983plbc}. We assume that the number of redundant symbols for masking defects and correcting random errors are $l$ and $r$, respectively. The total redundancy will be $l + r$, which is same as $n - k$ (where $n$ is the codeword size and $k$ is the message size). Note that the code rate is $k / n$.

The fact that the redundancy of PLBC can be divided into two parts leads to the problem of redundancy allocation. The objective is to find an optimal redundancy allocation between $l$ and $r$ in order to minimize the probability of decoding failure. Not surprisingly, the optimal redundancy allocation depends on the channel. If a channel exhibits only defects, we should allot all redundancy to masking defects and the optimal redundancy allocation will be $(l^{*}, r^{*})=(n-k, 0)$. Meanwhile, the optimal redundancy allocation for a channel with only random errors will be $(l^{*}, r^{*})=(0, n-k)$.

In this paper, the optimal redundancy allocation for general channels that exhibit both defects and random errors will be investigated. In addition, we will derive an upper bound on the probability of decoding failure. With this upper bound, we can readily obtain an estimate $(\widehat{l}, \widehat{r})$ for the optimal redundancy allocation $(l^{*}, r^{*})$. The estimate based on the upper bound matches the optimal redundancy allocation well.

The rest of the paper is as follows. Section II explains the channel model and PLBC. In section III, we will discuss the optimal redundancy allocation of PLBC. In Section IV, we will derive the upper bound on the probability of decoding failure and compare the optimal redundancy allocation with the estimate based on our upper bound. Section V concludes the paper.


\section{Partitioned Linear Block Codes \cite{heegard1983plbc, Kim2013}}
\subsection{Channel Model}
In \cite{heegard1983capacity, heegard1983plbc}, the channel model for memories with defects has been introduced. The model assumes both stuck-at defects and additive random errors. In this paper, we will use the notation of \cite{heegard1983plbc, Kim2013}.

Let $q$ be a power of a prime and $F_q$ be the Galois field with $q$ elements. Let $F_q^n$ denote the set of all $n$-tuples over $F_q$. Define an additional variable ``$\lambda$'', $\widetilde{F}̃_q = F_q\cup\lambda$ and define the ``$\circ$'' operator $\circ:F_q \times \widetilde{F}̃_q \rightarrow F_q$ by
\begin{equation}\label{eq:circ_operator}
x \circ s =
\begin{cases}
x, & \text{if } s = \lambda ; \\
s, & \text{if } s \ne \lambda.
\end{cases}
\end{equation}
An $n$-cell memory with defects and random errors is modeled by
\begin{equation}\label{eq:channel_model}
\mathbf{y} = (\mathbf{x} \circ \mathbf{s}) + \mathbf{z}
\end{equation}
where $\mathbf{x}$ is the vector to be stored, $\mathbf{z}$ is the random error vector and $\mathbf{s}$ is the defect vector. The addition ``$+$'' is defined over the field $F_q$ and both $+$ and $\circ$ operate on the vectors component wise.

The number of defects $u$ is equal to the number of non-$\lambda$ components in $\mathbf{s}$, and the number of random errors is defined by $t=\|\mathbf{z}\|$ where $\|\cdot\|$ is the Hamming weight of the vector.

A stochastic model for the generation of defects and random errors in a memory cell is obtained by assigning probabilities to defects and random error events. The $(\varepsilon, p)$ $q$-symmetric discrete memoryless memory cell ($q$-SDMMC) is modeled by $Y = (X \circ S) + Z$, where $X,Y,Z \in F_q, S \in \widetilde{F}̃_q$,
\begin{equation}\label{eq:random_channel_model}
\begin{aligned}
P(S=s)& =
\begin{cases}
1-\varepsilon,  & s = \lambda; \\
\frac{\varepsilon}{q}, &  s \ne \lambda,
\end{cases}
\\
P(Z=z|S=\lambda) &=
\begin{cases}
1-p, & z = 0; \\
\frac{p}{q - 1}, & z \ne 0.
\end{cases}
\end{aligned}
\end{equation}
Fig. 1 illustrates the channel model for memories with defects when $q=2$.

\begin{figure}[!t]
   \centering
   \includegraphics[width=0.30\textwidth]{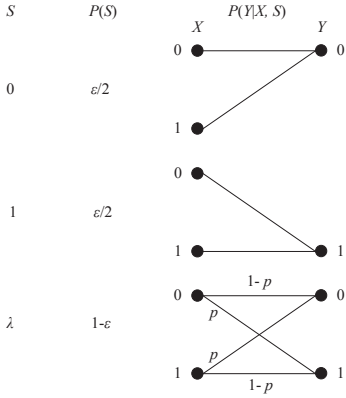}
   \caption{Channel model for binary memories with defects.}
   \label{fig:channel_model}
\end{figure}

\subsection{Partitioned Linear Block Codes}
In \cite{heegard1983plbc}, Heegard proposed the $[n,k,l]$ PLBC which is a pair of linear subspaces $\mathcal{C}_1 \subset F_q^n$ and $\mathcal{C}_0 \subset F_q^n$ of dimension $k$ and $l$ such that $\mathcal{C}_1 \cap \mathcal{C}_0 =\{ \mathbf{0}\}$. Then the direct sum
\begin{equation}\label{eq:direct_sum}
\mathcal{C} \triangleq \mathcal{C}_1 + \mathcal{C}_0 = \{ \mathbf{c} = \mathbf{c}_1 + \mathbf{c}_0 | \mathbf{c}_1 \in \mathcal{C}_1 , \mathbf{c}_0 \in \mathcal{C}_0 \}
\end{equation}
is an $[n,k+l]$ linear block code (LBC) with a generator matrix $G=[ G_1^T \quad G_0^T ]^T$ where $G_1$ generates $\mathcal{C}_1$ and $G_0$ generates $\mathcal{C}_0$ (superscript $T$ denotes transpose). The parity check matrix $H$ is $r \times n$ matrix with $k+l+r=n$. A message inverse matrix $\widetilde{G}̃_1$ is defined as $k \times n$ matrix with $G_1 \widetilde{G}̃_1^{T}=I_k$ (the $k$-dimensional identity matrix), and $G_0 \widetilde{G}̃_1^{T} =0_{l,k}$ (the $l \times k$ zero matrix) \cite{heegard1983plbc}.

The encoding and decoding of PLBC are as follows \cite{heegard1983plbc}.

\emph{Encoding:} To encode a message $\mathbf{w} \in F_q^k$ into a codeword $\mathbf{c}=\mathbf{w} G_1 + \mathbf{d} G_0$ where $\mathbf{d} \in F_q^l$ is chosen to minimize $ \|(\mathbf{c} \circ \mathbf{s}) - \mathbf{c} \|$.

\emph{Decoding:} Receive $\mathbf{y} = ( \mathbf{c} \circ \mathbf{s} ) + \mathbf{z}$. Compute the syndrome $\mathbf{v} = \mathbf{y} H^T$ and choose $\widehat{\mathbf{z}} \in F_q^n$ which minimizes $\| \mathbf{z} \|$ subject to $\mathbf{z} H^T = \mathbf{v}$. Then $\widehat{\mathbf{w}} = \widehat{\mathbf{c}} \widetilde{G}_1^T$ where $\widehat{\mathbf{c}}= \mathbf{y} - \widehat{\mathbf{z}}$.

The encoding of PLBC requires two generator matrices, namely, $G_0$ and $G_1$. First, $G_1$ encodes a message $\mathbf{w}$ for correcting random errors. 
Next, $G_0$ is used to mask defects by $\mathbf{d}G_0$ in order to minimize $ \|(\mathbf{c} \circ \mathbf{s}) - \mathbf{c} \|$. The redundancy for masking defects is $l$ and the redundancy for correcting random errors is $r$. Let $(l, r)$ denote the redundancy of $[n, k, l]$ PLBC. The total redundancy is $l+r = n-k$.

A pair of minimum distances $(d_0 , d_1)$ of an $[n,k,l]$ PLBC are given by
\begin{align}
d_0 &= \underset{
\substack{
\mathbf{c} \ne \mathbf{0} \\
\mathbf{c} G_0^T = \mathbf{0}
}}
{\text{min }} \|\mathbf{c}\|, \label{eq:min_d0}
\\
d_1 &= \underset{
\substack{
\mathbf{c} \widetilde{G}_1^T \ne \mathbf{0} \\
\mathbf{c} H^T = \mathbf{0}
}}
{\text{min }} \|\mathbf{c}\| \label{eq:min_d1}
\end{align}
where $d_1$ is greater than or equal to the minimum distance of the $[n,k+l]$ LBC with parity check matrix $H$, while $d_0$ is the minimum distance of the $[n,k+r]$ LBC with the parity check matrix $G_0$ \cite{heegard1983plbc}. Note that this $[n,k+r]$ LBC uses $G_0$ as a parity check matrix instead of a generator matrix.

\begin{theorem} \emph{\cite{heegard1983plbc}:} \label{th:1}
An $[n,k,l]$ PLBC with minimum distances $\left(d_0, d_1 \right)$ is a $u$-defect, $t$-error correcting code if
\begin{equation*}
u < d_0 \text{ and } 2t < d_1
\end{equation*}
or
\begin{equation*}
u \ge d_0 \text{ and } 2(t + u -(d_0 - 1)) < d_{1}.
\end{equation*}
\end{theorem}

If $u < d_0$, all defects will be successfully masked and $\|(\mathbf{c} \circ \mathbf{s})-\mathbf{c} \|=0$. Otherwise, it may be that $\|(\mathbf{c} \circ \mathbf{s})-\mathbf{c} \| \ne 0$ which results in masking failure. When $d_0 - 1$ defects among $u$ defects have been masked, the number of unmasked defects is $u - (d_0 -1)$. These unmasked defects will be regarded as random errors in the decoder.

Note that the masking succeeds in the encoder if and only if $\mathbf{c} \circ \mathbf{s} = \mathbf{c}$. In addition, $\widehat{\mathbf{w}} = \mathbf{w}$ means the decoding success.

\subsection{Two-step Encoding Scheme}
The encoding of PLBC includes an implicit optimization problem which can be formulated as follows \cite{heegard1983plbc, lastras2010algorithm, Hwang2011}.
\begin{equation}\label{eq:optimization_problem_encoding}
\begin{aligned}
\mathbf{d}^*  & =  \underset{\mathbf{d}}{\text{argmin }} \left\|\mathbf{d} G_0^{\Psi_u} + \mathbf{w} G_1^{\Psi_u} - \mathbf{s}^{\Psi_u} \right\|  \\
&= \underset{\mathbf{d}}{\text{argmin }} \left\|\mathbf{d} G_0^{\Psi_u} + \mathbf{b}^{\Psi_u} \right\|
\end{aligned}
\end{equation}
where $\Psi_u=\left[i_1,\cdots,i_u \right]$ indicates the locations of $u$ defects and $\mathbf{b}^{\Psi_u} = \mathbf{w} G_1^{\Psi_u} - \mathbf{s}^{\Psi_u}$. We use the notation of $s^{\Psi_u}=[s_{i_1},\cdots,s_{i_u}]$, $G_0^{\Psi_u}=[\mathbf{g}_{0,i_1},\cdots,\mathbf{g}_{0,i_u}]$, and $G_1^{\Psi_u}=[\mathbf{g}_{1,i_1 },\cdots,\mathbf{g}_{1,i_u}]$ where $\mathbf{g}_{0,i}$ and $\mathbf{g}_{1,i}$ are the $i$-th columns of $G_0$ and $G_1$ respectively. Then, $\|\mathbf{d} G_0^{\Psi_u} + \mathbf{b}^{\Psi_u} \|$ is the number of unmasked defects.

\begin{table}[t]
\renewcommand{\arraystretch}{1.2}
\caption{Two-Step Encoding Scheme}
\label{tab:twostep}
\centering
\begin{tabular}{p{8cm}}
\hline
Step 1:
\begin{itemize}
\item Try to solve \eqref{eq:linear_equation}.
    \begin{itemize}
        \item If $u < d_0$, a solution $\mathbf{d}$ to \eqref{eq:linear_equation} always exists and go to end.
        \item If $u \ge d_0$, a solution $\mathbf{d}$ to \eqref{eq:linear_equation} exists so long as \eqref{eq:rank_condition} holds.
            \begin{itemize}
                \item If $\mathbf{d}$ exists, go to end.
                \item Otherwise, go to step 2.
            \end{itemize}
    \end{itemize}
\end{itemize}
Step 2:
\begin{itemize}
\item Choose $d_0 - 1$ locations among $u$ defects and define $\Psi_{d_0-1}=\left[i_1,\cdots,i_{d_0-1} \right]$.
\item Solve $\mathbf{d} G_0^{\Psi_{d_0-1}}=\mathbf{b}^{\Psi_{d_0-1}}$ instead of \eqref{eq:linear_equation}.
\end{itemize}
End
\\ \hline
\end{tabular}
\end{table}

In order to mask all defects, we should have a solution $\mathbf{d}$ satisfying
\begin{equation} \label{eq:linear_equation}
\mathbf{d} G_0^{\Psi_u} =\mathbf{b}^{\Psi_u}.
\end{equation}
It is true that \eqref{eq:linear_equation} has at least one solution if and only if
\begin{equation} \label{eq:rank_condition}
\rank \left( \left(G_0^{\Psi_u} \right) ^T\right)= \rank \left( \left(G_0^{\Psi_u}\right)^T  \left| \left(\mathbf{b}^{\Psi_u}\right)^T\right. \right)
\end{equation}
where $ ( (G_0^{\Psi_u} )^T  |  (\mathbf{b}^{\Psi_u} )^T  )$ is the augmented matrix \cite{Friedberg1997}.

If $u<d_{0}$, $\rank\left(G_0^{\Psi_u}\right)$ is always $u$ by \eqref{eq:min_d0}. Therefore, \eqref{eq:rank_condition} holds and a solution $\mathbf{d}$ satisfying \eqref{eq:linear_equation} exists. Gaussian elimination or some other solution methods for linear equations can be used to solve \eqref{eq:linear_equation}.

However, if $u \ge d_0$, the optimal solution of \eqref{eq:optimization_problem_encoding} may fail to mask all defects. In addition, the computational complexity for solving the optimization problem is exponential, which is impractical as $l$ increases \cite{Hwang2011}.

In \cite{heegard1983plbc}, a modified formulation of \eqref{eq:optimization_problem_encoding} was described, which chooses only $\min \left( d_0-1, u \right)$ locations among $u$ defects instead of solving the optimization problem. Then, a solution of the modified formulation exists. This scheme achieves the $u$-defect, $t$-error correcting code of Theorem \ref{th:1} \cite{heegard1983plbc}. We call it \emph{one-step encoding scheme}.

In \cite{Kim2013}, the \emph{two-step encoding scheme} has been proposed, which can mask more defects the than one-step encoding scheme. The computational complexity of the two-step encoding scheme is comparable to that of the one-step encoding scheme. The two-step encoding scheme is summarized in Table \ref{tab:twostep}. We will use this two-step encoding scheme for encoding of PLBC.

\section{Optimal Redundancy Allocation}

\begin{table}[t]
\renewcommand{\arraystretch}{1.3}
\caption{Channels with the Same $C_{\min}$}
\label{tab:channel}
\centering
{\small
\hfill{}
\begin{tabular}{|c|c|c|c|c|}
\hline
Channel & {$p$}                & {$\varepsilon$}      & {$C_{\min}$} & {$C_{\max}$}  \\ \hline \hline
1       & $4.0 \times 10^{-3}$ & 0                    & 0.9624       & 0.9624        \\ \hline
2       & $3.0 \times 10^{-3}$ & $2.0 \times 10^{-3}$ & 0.9624       & 0.9686        \\ \hline
3       & $2.5 \times 10^{-3}$ & $3.0 \times 10^{-3}$ & 0.9624       & 0.9719        \\ \hline
4       & $2.0 \times 10^{-3}$ & $4.0 \times 10^{-3}$ & 0.9624       & 0.9753        \\ \hline
5       & $1.0 \times 10^{-3}$ & $6.0 \times 10^{-3}$ & 0.9624       & 0.9827        \\ \hline
6       & $5.0 \times 10^{-4}$ & $7.0 \times 10^{-3}$ & 0.9624       & 0.9868        \\ \hline
7       &  0                   & $8.0 \times 10^{-3}$ & 0.9624       & 0.9920        \\ \hline
\end{tabular}}
\hfill{}
\end{table}

\begin{table}[t]
\renewcommand{\arraystretch}{1.3}
\caption{All Possible Redundancy Allocation Candidates of $\left[ n = 1023, k=923, l \right]$ PBCH Codes}
\label{tab:PLBC}
\centering
{\small
\begin{tabular}{|c|c|c|c|c|c|}
\hline
Code & {$l$} & {$r$} & {$d_0$} & {$d_1$} & Notes   \\ \hline \hline
0 & 0 & 100 & 0 & 21 & only correcting random errors \\ \hline
1 & 10 & 90 & 3 & 19 &\\ \hline
2 & 20 & 80 & 5 & 17 &\\ \hline
3 & 30 & 70 & 7 & 15 &\\ \hline
4 & 40 & 60 & 9 & 13 &\\ \hline
5 & 50 & 50 & 11 & 11 & \\ \hline
6 & 60 & 40 & 13 & 9 &\\ \hline
7 & 70 & 30 & 15 & 7 &\\ \hline
8 & 80 & 20 & 17 & 5 &\\ \hline
9 & 90 & 10 & 19 & 3 &\\ \hline
10& 100 & 0 & 21 & 0 & only masking defects\\ \hline
\end{tabular}}
\end{table}

In order to minimize the probability of decoding failure of PLBC, we have to find the optimal redundancy allocation $(l^*, r^*)$. This problem can be formulated as follows.
\begin{equation}\label{eq:optimization_problem}
\begin{aligned}
(l^*, r^*)  = \; & \underset{(l, r)}{\text{argmin}} & & P(\text{decoding failure}) \\
& \text{subject to} & & l+r = n - k
\end{aligned}
\end{equation}
$(l^*, r^*)$ depends on the parameters of the given channel such as $\varepsilon$ and $p$. Without an expression for $P(\text{decoding failure})$ as a function of $\left(l, r\right)$, this optimization problem cannot be solved. Unfortunately, it is difficult to obtain the exact mathematical expression for $P(\text{decoding failure})$. Therefore, we obtained $(l^*, r^*)$ via simulation of the given channel and the given PLBC.

We will consider the channels of Table \ref{tab:channel}. All channels of Table \ref{tab:channel} are chosen to have the same $C_{\min}$, which is the channel capacity when neither the encoder nor the decoder knows the defect information \cite{heegard1983capacity}. $C_{\min}$ is given by
\begin{equation}\label{eq:capacity_min}
C_{\min} = 1 - h\left( \left(1 - \varepsilon \right) p + \frac{\varepsilon}{2} \right)
\end{equation}
where $h\left( x\right) = -x \log_2 x - \left(1-x\right) \log_2\left(1-x\right)$. Note that \eqref{eq:capacity_min} equals the capacity of a binary symmetric channel (BSC) with parameter $\widetilde{p} = \left(1 - \varepsilon \right) p + \frac{\varepsilon}{2}$. All channels of Table \ref{tab:channel} have the same $\widetilde{p}\cong4.0 \times 10^{-3}$. If either the encoder or the decoder knows the information of defects, the maximum capacity can be achieved \cite{heegard1983capacity}. The capacity is given by
\begin{equation}\label{eq:capacity_max}
C_{\max} = \left(1 - \varepsilon \right) \left( 1 - h \left(p \right) \right).
\end{equation}
Each channel of Table \ref{tab:channel} has different $C_{\max}$. As $\varepsilon$ increases, we can obtain more information about the channel, which results in the increase of $C_{\max}$.

For these channels of Table \ref{tab:channel}, we apply several kinds of  $\left[ n = 1023, k=923, l \right]$ PLBC. With the fixed code rate of $\frac{k}{n} = \frac{923}{1023}$, the redundancy allocation between $\left(l, r\right)$ will be varied. All possible redundancy allocation candidates of the partitioned Bose-Chaudhuri-Hocquenghem (PBCH) code are presented in Table \ref{tab:PLBC}. The PBCH code is a special class of PLBC and its generator matrices and minimum distances can be designed by a similar method of standard BCH codes \cite{heegard1983plbc}.

Fig. 2 shows the simulation results for the channels of Table \ref{tab:channel}. The optimal redundancy allocation for channel 1 (only random errors) will be $\left(l^*=0, r^*=n-k=100 \right)$, i.e., we have to allot all redundancy for correcting random errors, which is equivalent to the standard BCH code. The more defects a channel has, the larger $l$ is expected to be for the optimal redundancy allocation. Eventually, the optimal redundancy for channel 7 will be $(l^* = n-k=100, r^*=0)$ though the simulation result of channel 7 is incomplete because of its impractical computational complexity. The optimal $l^*$ for all channels of Table \ref{tab:channel} are presented in the second column of Table \ref{tab:redundancy}. The optimal $r^*$ can be obtained by $r^* = n - k - l^*$.

\begin{figure}[t]
   \centering
   \includegraphics[width=0.4\textwidth]{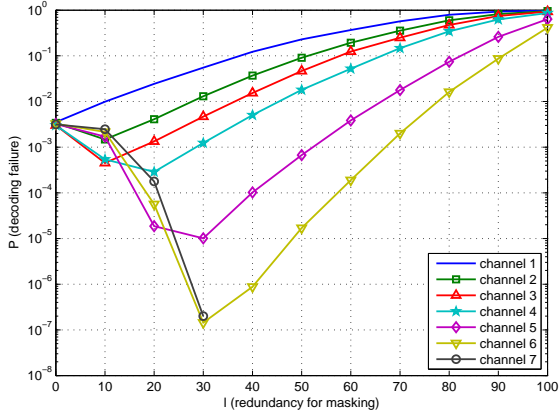}
   \caption{$P(\text{decoding failure})$ of channels in Table \ref{tab:channel}.}
   \label{fig:decoding_failure}
\end{figure}

\begin{table}[t]
\renewcommand{\arraystretch}{1.3}
\caption{Optimal Redundancy Allocation $l^*$ and its Estimate $\widehat{l}$ by Upper Bound for $\left[n=1023, k=923, l \right]$ PBCH Codes}
\label{tab:redundancy}
\centering
{\small
\begin{tabular}{|c|c|c|c|}
\hline
Channel & {$l^*$} & {$\widehat{l}$} & $| l^* - \widehat{l} |$    \\ \hline \hline
1 & 0  & 0  & 0  \\ \hline
2 & 10 & 10 & 0  \\ \hline
3 & 10 & 20 & 10 \\ \hline
4 & 20 & 20 & 0  \\ \hline
5 & 30 & 30 & 0  \\ \hline
6 & 30 & 30 & 0  \\ \hline
7 & 100& 100& 0 \\ \hline
\end{tabular}}
\end{table}

In addition, it is worth mentioning that $P(\text{decoding failure})$ for $(l^*, r^*)$ improves as $\varepsilon$ increases in Fig. 2. The reason is that the PLBC can exploit more information of defects, which has been indicated by $C_{\max}$ of Table \ref{tab:channel}. Note that $P(\text{decoding failure})$ for all channels will be same if we use the redundancy allocation $\left(l=0, r=n-k\right)$ when the encoder does not use the information of defects.

To find the optimal redundancy allocation $(l^*, r^*)$ by simulation requires significant computations. Therefore, we will also investigate using the upper bound on $P(\text{decoding failure})$ instead of the simulation for estimating the optimal redundancy allocation.

\section{Upper Bound on $P(\text{Decoding Failure})$}

In this section, the upper bound on $P(\text{decoding failure})$ will be derived. For convenience, we will define three random variables of $D$, $M$, and $U$.
\begin{equation}\label{eq:random_variable}
\begin{aligned}
D& =
\begin{cases}
0,  & \widehat{\mathbf{w}} \ne \mathbf{w}\text{ (decoding failure)}; \\
1,  & \widehat{\mathbf{w}} = \mathbf{w}\text{ (decoding success)},
\end{cases}
\\
M &=
\begin{cases}
0, & \mathbf{c} \circ \mathbf{s} \ne \mathbf{c}\text{ (masking failure)}; \\
1, & \mathbf{c} \circ \mathbf{s} = \mathbf{c}\text{ (masking success)}
\end{cases}
\end{aligned}
\end{equation}
In addition, $U$ represents the number of defects per codeword of PLBC. The probability of decoding failure $P(D=0)$ will be given by
\begin{equation}\label{eq:decoding_failure}
P(D=0) = P(M=0, D=0) + P(M=1, D=0).
\end{equation}

First, we will derive the upper bound on $P(M=0, D=0)$. By the chain rule, $P(M=0, D=0)$ is given by
\begin{IEEEeqnarray}{rCl}
{P(M=0, D=0)} &=& \sum_{u=1}^{n} { \left\{ P(U=u) \cdot P(M=0|U=u) \right.} \nonumber \\
&   & \quad \left. \cdot P(D=0|M=0,U=u) \right\}.
\label{eq:P_M0D0}
\end{IEEEeqnarray}
Note that we do not need to consider $u=0$ since the masking always succeeds for $u=0$.

By \eqref{eq:random_channel_model}, we can assume that $U$ is a binomial random variable. Therefore, $P(U=u)$ is given by
\begin{equation}\label{eq:P_Uu}
P(U=u)= \binom{n}{u} \varepsilon^{u} \left( 1 - \varepsilon \right)^{n-u}, \text{   } 0 \le u \le n.
\end{equation}
In \cite{Kim2013}, the following upper bound on $P(M=0 |U=u)$ was derived. 
\begin{equation}\label{eq:P_M0Uu}
P(M=0|U=u) \le  \frac{\sum_{w=d_0}^{u}{A_w \binom{n-w}{u-w}}}{\binom{n}{u}}
\end{equation}
where $A_w$ is the number of codewords of weight $w$ in the LBC with the parity check matrix $G_0$. Note that $A_w=0$ for $0 < w < d_0$ by \eqref{eq:min_d0}.

Also, $P(D=0 | M=0, U=u)$ is given by
\begin{IEEEeqnarray}{rCl}
{P(D=0|M=0, U=u)}
& = & P\left( \{ u - \left( d_0 - 1 \right) \} + t > t_1 \right) \nonumber \\
& = & P\left( t \ge t_1 + d_0 - u \right)\label{eq:P_D0M0Uu1}
\end{IEEEeqnarray}
where $u - \left( d_0 - 1 \right)$ represents the number of unmasked defects and $t$ is the number of random errors. In addition, $t_1 = \lfloor \frac{d_1 - 1}{2} \rfloor$ (where $\lfloor x \rfloor$  is the largest integer not greater than $x$) is the error correcting capability of $\mathcal{C}_1$. Since the number of random errors can be modeled by the binomial random variable by \eqref{eq:random_channel_model}, $P(D=0 | M=0, U=u)$ is given by
\begin{IEEEeqnarray}{rCl}
\IEEEeqnarraymulticol{3}{l}
{P\left( D=0 | M=0, U=u  \right) }\nonumber \\
\quad
& = &\sum_{t=t_1 + d_0 - u}^{n - u}{\binom{n-u}{t} p^{t} \left( 1 - p \right)^{n-u -t}}.\label{eq:P_D0M0Uu2}
\end{IEEEeqnarray}

By substituting \eqref{eq:P_Uu}, \eqref{eq:P_M0Uu} and \eqref{eq:P_D0M0Uu2} into \eqref{eq:P_M0D0}, the upper bound on $P(M=0, D=0)$ is given by
\begin{IEEEeqnarray}{rCl}
\IEEEeqnarraymulticol{3}{l}
{P(M=0, D=0)}\nonumber \\
\quad
& \le & \sum_{u=d_0}^{n} \left\{ {\binom{n}{u} \varepsilon^{u} \left( 1 - \varepsilon \right)^{n-u}} \cdot \frac{\sum_{w=d_0}^{u}{A_w \binom{n-w}{u-w}}}{\binom{n}{u}} \right.  \nonumber \\
&   &  \left. \cdot \sum_{t=t_1 + d_0 - u}^{n - u}{\binom{n-u}{t} p^{t} \left( 1 - p \right)^{n-u -t}} \right\}.
\label{eq:P_M0D0_ub}
\end{IEEEeqnarray}

Now, we will derive the upper bound on $P(M=1, D=0)$ in \eqref{eq:decoding_failure}. By the chain rule,
\begin{multline}
{P(M=1, D=0)} = \sum_{u=0}^{n} { \left\{ P(U=u) \cdot P(M=1|U=u) \right.}  \\
\left. \cdot P(D=0|M=1,U=u) \right\}
\label{eq:P_M1D0}
\end{multline}
where $P(U=u)$ was given by \eqref{eq:P_Uu} and $P(M=1|U=u) \le 1$. Also, $P(D=0 | M=1, U=u)$ is given by
\begin{IEEEeqnarray}{rCl}
\IEEEeqnarraymulticol{3}{l}
{P(D=0|M=1, U=u) = P(t > t_1)}\nonumber \\
\quad \quad
& = & \sum_{t=t_1 + 1}^{n-u}{\binom{n-u}{t}p^t (1-p)^{n-u-t}}.
\label{eq:P_D0M1Uu}
\end{IEEEeqnarray}

By substituting \eqref{eq:P_Uu}, \eqref{eq:P_D0M1Uu}, and $P(M=1|U=u) \le 1$ into \eqref{eq:P_M1D0}, the upper bound on $P(M=1, D=0)$ is given by
\begin{multline}
{P(M=1, D=0)} \le \sum_{u=0}^{n} \Biggl\{ {\binom{n}{u} \varepsilon^{u} \left( 1 - \varepsilon \right)^{n-u}} \cdot \Biggr.   \\
  \Biggr. \cdot \sum_{t=t_1 + 1}^{n-u}{\binom{n-u}{t}p^t (1-p)^{n-u-t}} \Biggr\}.
\label{eq:P_M1D0_ub}
\end{multline}

By substituting \eqref{eq:P_M0D0_ub} and \eqref{eq:P_M1D0_ub} into \eqref{eq:decoding_failure}, $P(D=0)$ is given by
\begin{IEEEeqnarray}{rCl}
\IEEEeqnarraymulticol{3}{l}
{P(D=0)}\nonumber \\
\quad
& \le & \sum_{u=d_0}^{n} \left\{ {\binom{n}{u} \varepsilon^{u} \left( 1 - \varepsilon \right)^{n-u}} \cdot \frac{\sum_{w=d_0}^{u}{A_w \binom{n-w}{u-w}}}{\binom{n}{u}} \right.  \nonumber \\
&   &  \left. \cdot \sum_{t=t_1 + d_0 - u}^{n - u}{\binom{n-u}{t} p^{t} \left( 1 - p \right)^{n-u -t}} \right\} \nonumber \\
& + & \sum_{u=0}^{n} \Biggl\{ {\binom{n}{u} \varepsilon^{u} \left( 1 - \varepsilon \right)^{n-u}} \cdot \Biggr.  \nonumber \\
&   & \Biggr. \cdot \sum_{t=t_1 + 1}^{n-u}{\binom{n-u}{t}p^t (1-p)^{n-u-t}} \Biggr\}.
\label{eq:P_D0_ub}
\end{IEEEeqnarray}

For $\varepsilon = 0$, we can claim that $P(M=0, D=0)=0$ since the upper bound on $P(M=0, D=0)$ becomes zero by \eqref{eq:P_M0D0_ub}. In addition, the terms of \eqref{eq:P_M1D0_ub} also become zero for $u \ge 1$. Therefore, \eqref{eq:P_D0_ub} will be changed into
\begin{equation}\label{eq:P_D0_ub_e0}
P(D=0) \le \sum_{t=t_1 + 1}^{n}{\binom{n}{t}p^t (1-p)^{n-t}}.
\end{equation}

For $l=0$, the PLBC is same as the standard error control coding which does not know the information of defects. Therefore, $P(D=0)$ will be as follows instead of \eqref{eq:P_D0_ub}.
\begin{equation}\label{eq:P_D0_ub_l0}
P(D=0) \le \sum_{t=t_1 + 1}^{n}{\binom{n}{t} \widetilde{p}^t \left( 1-\widetilde{p} \right)^{n-t}}
\end{equation}

By \eqref{eq:P_D0_ub}, \eqref{eq:P_D0_ub_e0} and \eqref{eq:P_D0_ub_l0}, we can readily obtain the upper bound on the probability of decoding failure for the given channel and all possible redundancy allocation candidates such as Table \ref{tab:PLBC}. Then, we can choose the redundancy allocation minimizing the upper bound instead of the probability of decoding failure. The redundancy allocation that minimizes the upper bound is the estimate of the optimal redundancy allocation.

The estimates $\widehat{l}$ for all channels of Table \ref{tab:channel} are presented in the third column of Table \ref{tab:redundancy}. Note that $\widehat{r} = n - k - \widehat{l}$. The estimates by the upper bound match the optimal redundancy allocation well. The only exception occurs in channel 3, and the difference of $l$ is only 10 bits, which is the smallest difference among all possible redundancy allocation candidates of Table \ref{tab:PLBC}.


\section{Conclusion}

The redundancy allocation of PLBC was discussed and the optimal redundancy allocation was investigated via simulation. In addition, we derived an upper bound on decoding failure probability of PLBC for general channels, and used it for estimating the optimal redundancy allocation. The estimated redundancy allocation is very similar to the optimal redundancy allocation whereas it requires much less computation than simulations required for determining the optimal redundancy allocation.


%




\end{document}